\documentclass[11pt,a4paper]{JHEP3}
\usepackage{amssymb}
\usepackage{euscript}

\def\a{{\alpha}}
\def\b{{\beta}}
\def\d{\partial}

\def\e{\epsilon}

\def\0{\nonumber}

\def\x{{\bf x}}
\def\det{{\rm Det}}

\newcommand\ES{\EuScript{S}}

\newcommand\T{\EuScript{T}}

\newcommand\X{\EuScript{X}}
\newcommand\V{\EuScript{V}}

\newcommand\EP{\EuScript{P}}

\newcommand\I{\mathbb{I}}
\newcommand\ee{\end{eqnarray}}	 	
\newcommand\be{\begin{eqnarray}}
\newcommand\ba{\begin{array}}			
\newcommand\ea{\end{array}}
\newcommand\eeq{\end{equation}}	 	
\newcommand\beq{\begin{equation}}

\preprint{SISSA/50/02/EP\\\tt hep-th/0207044}

\title{Vacuum String Field Theory ancestors of the GMS solitons}

\author{ L.Bonora, D.Mamone, M.Salizzoni\\
International School for Advanced Studies (SISSA/ISAS)\\
Via Beirut 2--4, 34014 Trieste, Italy, and INFN, Sezione di
Trieste\\
E-mail:   \email{bonora@sissa.it}, \email{mamone@sissa.it},
\email{sali@sissa.it} }

\abstract{We define a sequence of VSFT D--branes whose low energy limit
leads exactly to a corresponding sequence of GMS solitons. The D--branes are
defined by acting on a fixed VSFT lump with operators defined by means of
Laguerre polynomials whose argument is quadratic in the string creation operators.
The states obtained in this way form an algebra under the SFT star
product, which is isomorphic to a corresponding algebra of GMS solitons
under the Moyal product. In order to obtain a regularized field 
theory limit we embed the theory in a constant background $B$ field.}


\begin{document}
\section{Introduction}

The resemblance between Witten's star product \cite{W1} in
open string field theory (SFT) and the Moyal product in noncommutative
field theory is intriguing. There have been attempts to relate
them to each other. It has been shown first of all that they are
compatible: from the analysis of \cite{W2} and \cite{Sch}
it is clear that, if we restrict ourselves to the field theory regime
 in the presence of a constant background $B$ field,
the SFT $*$ product factorizes into the ordinary (Witten's) $*$ product
and the Moyal product.
On a different ground in \cite{bars,doug} it was suggested that
SFT can be formulated in terms of Moyal products, although in an
(unphysical) auxiliary space, see \cite{Bo2,Chen,belov,Chu}.

In this paper we would like to show that the similarity between
the two products is deeper that one may suspect. In the course
of the paper we will single out an interesting algebraic
structure for a family of solitonic solutions of the Vacuum String
Field Theory (VSFT), \cite{RSZ1,RSZ2}. This structure turns out to be
exactly isomorphic to a corresponding one for a
family of solitonic solutions in noncommutative field theory
found by Gopakumar, Minwalla and Strominger, \cite{GMS,Komaba}.

The classical solutions of Vacuum String Field Theory (VSFT) 
factorize into a ghost part and a matter part. The ghost factor 
is universal, while the matter may take different forms, but they 
all satisfy the projector equation:
\be
\Psi_m * \Psi_m \, =\, \Psi_m\label{proj}
\ee
Well--known solutions are the {\it squeezed or sliver state
solution} \cite{KP}, which is interpreted as the D25--brane 
\cite{RSZ2, Okawa}, and the analogous lower dimensional tachyonic lumps. 
Other solutions have been found in \cite{RSZ3,GT}.
It has been shown in \cite{BMS2} that, in the field theory limit,
the 23 dimensional lump solution can be identified with the 
simplest GMS soliton. However all the solutions that have 
been found so far reproduce,
in the field theory limit, only the first and second simplest cases
of the infinite series of (noncommutative) GMS solitons,
\cite{GMS,Komaba}.
Since the GMS solitons can be thought of as solutions of the field
theory limit of SFT, it is natural to imagine that there must exist
a full series of SFT solutions which has not been found so far.

In this paper, starting from the squeezed state, we
construct an infinite sequence of solutions to eq.(\ref{proj}),
denoted $|\Lambda_n\rangle$ for any natural number $n$.
$|\Lambda_n\rangle$ is generated by acting on a tachyonic lump
solution $|\Lambda_0\rangle$ with $(-\kappa)^nL_n(\x/\kappa)$, where $L_n$ is the $n$-th Laguerre polynomial, 
$\x$ is a quadratic expression in the string creation
operators, see below eqs.(\ref{x}, \ref{Lambdan}), and $\kappa$ is an arbitrary constant.  These states
satisfy the remarkable properties
\be
&& |\Lambda_n\rangle * |\Lambda_m\rangle = \delta_{n,m}
|\Lambda_n\rangle \label{nstarm}\\
&& \langle \Lambda_n |\Lambda_m\rangle = \delta_{n,m}
 \langle \Lambda_0 |\Lambda_0\rangle \label{nm}
\ee
Each $|\Lambda_n\rangle$ represents a D23--brane, parallel to all the 
others.
The field theory limit of $|\Lambda_n\rangle$ factors into
the sliver state (D25--brane) and the $n$-th GMS soliton.
The algebra (\ref{nstarm}) and the property (\ref{nm}) exactly
reflect isomorphic properties of the GMS solitons (in terms of Moyal
product). In other words, the GMS solitons are nothing but the relics
of the $|\Lambda_n\rangle$ D23--branes in the low energy limit.

In the following, to avoid the singularities of the field theory limit
of VSFT, \cite{MT}, we introduce a constant background $B$ 
field\footnote{One could implement the construction of this paper  
also without a $B$ field. However, in the latter case, in the
field theory limit one would have to 
introduce a regulator by hand, since, as shown in \cite{MT}, 
this limit is singular. With a nonvanishing background 
$B$ field we can avoid such an ad hoc procedure, as $B$ itself 
provides a natural regularization.}.
However it should be stressed that the proof of (\ref{nstarm},
\ref{nm}) hold for any value of $B$.

The paper is organized as follows. In the next section we collect
from the literature a series of results which are needed in the
following. In section 3 we define
the above solutions and prove eqs.(\ref{nstarm}, \ref{nm}).
In section 4 we find the field theory limit and recover the GMS
solitons. Section 5 is devoted to a discussion of the above 
mentioned isomorphism. Finally, Appendix A contains a discussion 
of the existence of our solutions, while Appendix B is devoted
to the $\a'\to 0$ limit.

\section{A collection of formulas and results}

To find in a natural way the field theory limit we introduce a
background $B$ field.
This problem was studied in \cite{sugino,KT,BMS,BMS2}. In the last two references
we found solutions to eq.(\ref{proj})
when a constant $B$ field is turned on along some space
directions. It will be enough to consider the simplest $B$ field
configuration, i.e. a $B$ field nonvanishing only in two space
directions, say the $24$--th and $25$--th ones. Let us denote
these directions with the Lorentz indices $\alpha$ and $\beta$.
Then, as is well--known \cite{SW}, in these two direction
we have a new effective metric $G_{\alpha\beta}$,
the open string metric, as well as an effective antisymmetric
parameter $\theta^{\alpha\beta}$. If we set
\be
B_{\a\b}= \left(\matrix {0&B\cr -B&0\cr}\right)\label{B}
\ee
with $B\geq 0$, they take the explicit form
\be
G_{\alpha\beta} = \sqrt{{\rm Det G}} \,\delta_{\a\b},\quad\quad
\theta^{\alpha\beta} =  -(2\pi \a')^2  B\epsilon^{\a\b} ,\quad\quad
{\rm Det G} = \left( 1+ (2\pi \a' B)^2\right)^2,  \label{Gtheta}
\ee
where $\epsilon^{\a\b}$ is the $2\times 2$ antisymmetric symbol
with $\epsilon^1{}_2=1$. $\a,\b$ indices are raised and lowered
using $G$.

The presence of the $B$ field modifies the three--string vertex
only in the 24-th and 25-th direction, which, in view of the D--brane
interpretation, we call the transverse ones.
After turning on the $B$--field the three--string vertex becomes
\be
|V_3 \rangle' =
|V_{3,\perp}\rangle ' \,\otimes\,|V_{3,\|}\rangle \label{split'}
\ee
$|V_{3,\|}\rangle$ is the same as in the ordinary case
(without $B$ field), while
\be
|V_{3,\perp}\rangle'= K_2\, e^{-E'}|\tilde 0\rangle_{123}\label{V3'}
\ee
with
\be
&&K_2= \frac {\sqrt{2\pi b^3}}{A^2 (4a^2+3)}({\rm Det} G)^{1/4},\label{K2}\\
&&E'= \frac 12 \sum_{r,s=1}^3 \sum_{M,N\geq 0} a_M^{(r)\a\dagger}
\V_{\a\b,MN}^{rs} a_N^{(s)\b\dagger}\label{E'}
\ee
We have introduced the indices $M=\{0,m\}, N=\{0,n\}$ and the
vacuum $|\tilde 0\rangle =
|0\rangle \otimes |\Omega_{b,\theta}\rangle$, where
$| \Omega_{b,\theta}\rangle$ is the vacuum with respect to the
oscillators
\be
a_0^{(r)\alpha} = \frac 12 \sqrt b \hat p^{(r)\alpha}
- i\frac {1}{\sqrt b} \hat x^{(r)\alpha},
\quad\quad
a_0^{(r)\alpha\dagger} = \frac 12 \sqrt b \hat p^{(r)\alpha} +
i\frac {1}{\sqrt b}\hat x^{(r)\alpha}, \label{osc}
\ee
where $\hat p^{(r)\alpha}, \hat x^{(r)\alpha}$ are the zero momentum
and position operator of the $r$--th string; i.e.
$a_0^\alpha|\Omega_{b,\theta}\rangle=0$.
It is understood that $p^{(r)\alpha} = G^{\a\b}p^{(r)}_\b$, and
\be
\big[a_M^{(r)\alpha},a_N^{(s)\beta\dagger}\big]=
G^{\alpha\beta}\delta_{MN}\delta^{rs}
\label{aNaN}
\ee
The coefficients $\V_{MN}^{\a\b,rs}$ are given by
\be
&&\V_{00}^{\a\b,rs} = G^{\a\b}\delta^{rs}- \frac {2A^{-1}b}{4a^2+3}
\left(G^{\a\b} \phi^{rs} -ia \epsilon^{\a\b}\chi^{rs}\right)
\label{VV00}\\
&&\V_{0n}^{\a\b,rs} = \frac {2A^{-1}\sqrt b}{4a^2+3}\sum_{t=1}^3
\left(G^{\a\b} \phi^{rt} -ia\epsilon^{\a\b}\chi^{rt}\right)
V_{0n}^{ts}\label{VV0n}\\
&&\V_{mn}^{\a\b,rs} = G^{\a\b}V_{mn}^{rs}-
\frac {2A^{-1}}{4a^2+3}\sum_{t,v=1}^3
V_{m0}^{rv}\left(G^{\a\b} \phi^{vt}
-ia \epsilon^{\a\b}\chi^{vt}\right)V_{0n}^{ts}\label{VVmn}
\ee
Here, by definition, $V_{0n}^{rs}=V_{n0}^{sr}$, and
\be
\phi^{rs}= \left(\matrix{1& -1/2& -1/2\cr
                    -1/2& 1& -1/2\cr
                     -1/2 &-1/2 &1}\right),\quad\quad
\chi^{rs}= \left(\matrix{0&1&-1\cr -1&0&1\cr 1&-1&0}\right)
\label{phichi}
\ee
These two matrices satisfy the algebra
\be
\chi^2 = - 2\phi,\quad\quad \phi\chi=\chi\phi = \frac 32 \chi,\quad\quad
\phi^2= \frac 32 \phi\label{chiphi}
\ee
Moreover, in (\ref{VVmn}), we have introduced the notation
\be
A = V_{00}+ \frac b2,  \quad\quad\quad
a = -\frac {\pi^2}A\, B,
\label{definitions}
\ee

Next we introduce the twist matrix $C'$ by $C'_{MN}= (-1)^M
\delta_{MN}$ and define
\be
\X^{rs} \equiv C' \V^{rs}, \quad\quad r,s=1,2,
\quad\quad \X^{11}\equiv \X\label{X}
\ee
These matrices commute
\be
[\X^{rs}, \X^{r's'}] =0 \label{commute}
\ee
Moreover we have the following properties, which mark a difference
with the $B=0$ case,
\be
C'\V^{rs}= \tilde \V^{sr}C' ,\quad\quad C'\X^{rs}= \tilde \X^{sr}C'
\label{CV}
\ee
where tilde denotes transposition with respect to the $\a,\b$
indices alone. Finally one can prove that
\be
&&\X+ \X^{12}+ \X^{21} = \I\0\\
&& \X^{12}\X^{21} = \X^2-\X\0\\
&& (\X^{12})^2+ (\X^{21})^2= \I- \X^2\0\\
&& (\X^{12})^3+ (\X^{21})^3 = 2 \X^3 -  3\X^2 +\I\label{Xpower}
\ee
In the matrix products of these identities, as well as throughout
the paper, the indices $\a,\b$ must be understood in alternating
up/down position: $\X^{\a}{}_{\b}$. For instance, in (\ref{Xpower})
$\I$ stands for $\delta^\a{}_\b\,\delta_{MN}$.

The lump solution we found in \cite{BMS} satisfies
$|\ES\rangle=|\ES\rangle * |\ES\rangle$ and can be written as
\be
|\ES\rangle \!&=&\! \left\{{\rm Det}(1-X)^{1/2}{\rm Det} (1+T)^{1/2}\right\}^{24}
{\rm exp}\left(-\frac 12 \eta_{\bar \mu\bar \nu}\sum_{m,n\geq 1}
a_m^{\bar \mu\dagger}S_{mn}a_n^{\bar \nu\dagger}\right)|0\rangle
\otimes\label{fullsol}\\
&& \frac {A^2 (3+4a^2)}{\sqrt{2 \pi b^3}({\rm Det}G)^{1/4}}
\left( {\rm Det}(\I -\X)^{1/2}{\rm Det}(\I + \T)^{1/2}\right)
{\rm exp}\left(-\frac 12 \sum_{M,N\geq 0}
a_M^{\a\dagger}\ES_{\a\b,MN}a_N^{\b\dagger}\right)
|\tilde 0 \rangle,\0
\ee
The quantities in the first line are defined in ref.\cite{RSZ2}
with $\bar\mu,\bar\nu=0,\ldots, 23$ denoting the parallel
directions to the lump. The matrix $\ES= C'\T $ is given by
\be
\T = \frac 1{2\X}\left( \I +\X - \sqrt{(\I + 3\X)(\I-\X)}
\right) \label{sol2}
\ee
This is a solution to the equation
\be
\X \T^2 - (\I + \X)\T + \X=0\label{equation}
\ee

Another ingredient we need in order to construct new solutions
starting from (\ref{fullsol}) are projectors similar to those
introduced in \cite{RSZ3}. They are defined only along the
transverse directions by
\be
\rho_1 \!&=&\! \frac 1{(\I +\T)(\I-\X)} \left[ \X^{12} (\I-\T\X)
+\T (\X^{21})^2\right]\label{rho1}\\
\rho_2 \!&=&\! \frac 1{(\I +\T)(\I-\X)} \left[ \X^{21} (\I-\T\X)
+\T (\X^{12})^2\right]\label{rho2}
\ee
They satisfy
\be
\rho_1^2 = \rho_1,\quad\quad \rho_2^2 = \rho_2, \quad\quad
\rho_1+\rho_2 = \I\label{proj12}
\ee
i.e. they project onto orthogonal subspaces. Moreover, if we
use the superscript $^T$ to denote transposition with respect to
the indices $N,M$ and $\a,\b$, we have
\be
\rho_1^T=\tilde\rho_1 = C'\rho_2 C',\quad\quad
 \rho_2^T=\tilde\rho_2 = C'\rho_1 C'.\label{rhorels}
\ee

With all these ingredients we can now move on, give a precise
definition of the $|\Lambda_n\rangle$ states and demonstrate the
properties announced in the introduction.

\section{The states $|\Lambda_n\rangle$ and their properties}

To define the states $|\Lambda_n\rangle$ we start from the lump
solution (\ref{fullsol}). I.e. we take $|\Lambda_0\rangle= |\ES\rangle$.
However, in the following, we will limit ourselves only to the transverse
part of it, the parallel one being universal and
irrelevant for our construction. We will denote the transverse part
by $|\ES_\perp\rangle$.

First we introduce two `vectors' $\xi=\{\xi_{N\a}\}$ and $\zeta =
\{\zeta_{N\a}\}$, which are chosen to satisfy the conditions
\be
\rho_1 \xi =0,\quad\quad \rho_2 \xi =\xi, \quad\quad
{\rm and}\quad \rho_1 \zeta =0,\quad\quad \rho_2\zeta=\zeta,
\label{xizeta}
\ee

Next we define
\be
\x = (a^\dagger \tau \xi)\, (a^\dagger C' \zeta)=
(a_N^{\a\dagger} \tau_\a{}^\beta \xi_{N\b})
(a_N^{\a\dagger}C'_{NM}\zeta_{M\a})   \label{x}
\ee
where $\tau$ is the matrix $\tau = \{ \tau_\a{}^\b\}=$
\mbox{ \small $\left(\begin{array}{cc}1 & 0 \\ 0 
& -1 \end{array}\right)$ },
and introduce the Laguerre polynomials $L_n(\x/\kappa)$.
The definition
of $|\Lambda_n\rangle$ is as follows
\be
|\Lambda_n\rangle = (-\kappa)^n L_n\Big(\frac{\x}{\kappa}\Big) |\ES_\perp\rangle\label{Lambdan}
\ee
where $\kappa$ is an arbitrary constant. Hermiticity requires that
\be
(a \tau \xi^*)(aC' \zeta^*) = (a\tau C'\xi)(a \zeta)\label{hermit}
\ee
and that $\kappa$ be a real constant. The superscript $^*$ denotes
complex conjugation.

Finally we impose that the two following conditions be satisfied
\be
\xi^T \tau\frac 1{\I-\T^2}\zeta =-1 ,\quad\quad
\xi^T \tau\frac {\T}{\I-\T^2}\zeta= -\kappa
\label{cond}
\ee

Compatibility of eqs.(\ref{xizeta},\ref{hermit},\ref{cond}) is discussed
in Appendix A. It is shown there that a solution to (\ref{hermit})
compatible with the other equations is 
\be
\zeta= \tau\xi^*.\label{unicum}
\ee
 
Before we set out to prove the properties of these states
$|\Lambda_n\rangle$, let us spend a few words to motivate their
definition. The definition (\ref{Lambdan}) is not, as one might suspect,
dictated in the first place by the similarity with the form of the GMS
solitons.
Rather it has been selected due to its apparently unique role
in the framework of Witten's star algebra.

In \cite{BMS2}, on the wake of \cite{RSZ3}, starting from the
(transverse) lump solution $|\ES_\perp\rangle$
we introduced a new lump solution
$|\EP_\perp\rangle = (\x - \kappa)
|\ES_\perp\rangle $. Imposing that $|\EP_\perp\rangle *
|\EP_\perp\rangle = |\EP_\perp\rangle$ and $|\EP_\perp\rangle *
|\ES_\perp\rangle =0$ and, moreover, that $\langle \EP_\perp|\EP_\perp\rangle =
\langle \ES_\perp|\ES_\perp\rangle $, we found 
the conditions (\ref{cond}).

The next most complicated state one is lead to try is of the form
\be
|\EP'\rangle =(\a + \b \x + \gamma \x^2) |\ES_\perp\rangle \label{try}
\ee
The conditions this state has to satisfy
 turn out to be more restrictive than for $|\EP\rangle$, but, nevertheless,
 are satisfied if, besides conditions (\ref{cond}), the following relations
hold
\be
-2(\a)^{1/2} = \b, \quad\quad  \gamma = \frac12
\ee
and then, putting $\a = \kappa$
\be
|\EP'\rangle =\Big(\kappa^2 - 2 \kappa\x + \frac 12 \x^2\Big) |\ES_\perp\rangle\label{try'}
\ee
The polynomial in the RHS is nothing but the second Laguerre
polynomial of $\x/\kappa$ multiplied by $\kappa^2$. We deduce from this that the Laguerre polynomials must
play a fundamental role in this problem and, as a consequence, put forward
the general ansatz (\ref{Lambdan}).

Proving the necessity of the conditions (\ref{cond}) for general $n$ is very cumbersome,
so we will limit ourselves to showing that these conditions are sufficient. However it
is instructive and rather easy to see, at least, that the second condition (\ref{cond}) is
necessary in general.
In fact, by requiring that the  state
$|\Lambda_n\rangle$  be orthogonal to the `ground state' $|\ES_\perp\rangle$, we get:
\be
|\Lambda_n\rangle *|\ES_\perp\rangle & = & (-\kappa)^n \sum_{j=0}^{\infty}
 \left(\matrix {n \cr j}\right) \frac{(-\x/\kappa)^j}{j!}
|\ES_\perp\rangle * |\ES_\perp\rangle \0 \\
 &=& (-\kappa)^n \sum_{j=0}^{\infty}
 \left(\matrix {n \cr j}\right)(\kappa)^{-j} \0\\
& \cdot & (\xi \tau C')_{l_1}^{\a_1}\ldots (\xi \tau C')_{l_j}^{\a_j}
\zeta_{k_1}^{\b_1}\ldots \zeta_{k_j}^{\b_j}
\frac {\d}{\d \mu_{l_1}^{\a_1}}\ldots \frac {\d}{\d \mu_{l_j}^{\a_j}}
\frac {\d}{\d \mu_{k_1}^{\b_1}}\ldots \frac {\d}{\d \mu_{k_j}^{\b_j}}\0\\
& \cdot &\exp\Big( - (\chi^T {\cal K}_{1})^{-1} \mu - \frac 12
\mu^T ({\cal V}{\cal K}^{-1})_{11}\mu \Big)
|\ES_\perp\rangle\Big\vert_{\mu=0} \0\\
&=&
(-\kappa)^n \sum_{j=0}^{\infty}
 \left(\matrix {n \cr j}\right)(\kappa)^{-j}
\left(\xi^T \tau\frac {\T}{\I-\T^2}\zeta \right)^j \0
|\ES\perp\rangle\\
&=&
(-\kappa)^n\left(1+\frac{1}{\kappa}\xi^T \tau\frac {\T}{\I-\T^2}\zeta\right)^n
|\ES_\perp\rangle =0
\ee
which is true for the choice $\kappa$ given by the second eq.(\ref{cond}).
In order to obtain this result we have followed ref.\cite{RSZ3} (see also \cite{BMS2}).

\subsection{Proof of eq.(1.2)}

The star product $|\Lambda_n\rangle * |\Lambda_{n'}\rangle$ can
be evaluated by using the explicit expression of the Laguerre
polynomials
\be
|\Lambda_n\rangle * |\Lambda_{n'}\rangle = \Big( (-\kappa)^n
\sum_{k=0}^n \left(\matrix {n\cr k\cr}\right)
\frac {(-\x/\kappa)^k}{k!}
|\ES_\perp\rangle \Big) * \Big( (-\kappa)^{n'}
\sum_{p=0}^{n'} \left(\matrix {n'\cr p\cr}\right)
\frac {(-\x/\kappa)^p}{p!}
|\ES_\perp\rangle \Big)\label{expanstar}
\ee
Therefore we need to compute $(\x^k |\ES_\perp\rangle) *
(\x^p |\ES_\perp\rangle)$. According to \cite{RSZ3}, this is given by
\be
(\x^k |\ES_\perp\rangle) * (\x^p |\ES_\perp\rangle)\!&=&\!
(\xi \tau C')_{l_1}^{\a_1}\ldots (\xi \tau C')_{l_k}^{\a_k}
\zeta_{j_1}^{\b_1}\ldots \zeta_{j_k}^{\b_k}
\frac {\d}{\d \mu_{l_1}^{\a_1}}\ldots \frac {\d}{\d \mu_{l_k}^{\a_k}}
\frac {\d}{\d \mu_{j_1}^{\b_1}}\ldots \frac {\d}{\d \mu_{j_k}^{\b_k}}
\0\\
&\cdot& (\xi \tau C')_{\bar l_1}^{\bar \a_1}\ldots
(\xi \tau C')_{\bar l_p}^{\bar\a_p}
\zeta_{\bar j_1}^{\bar\b_1}\ldots \zeta_{\bar j_p}^{\bar \b_p}
\frac {\d}{\d \bar\mu_{\bar l_1}^{\bar \a_1}}\ldots
\frac {\d}{\d \bar\mu_{\bar l_p}^{\bar \a_p}}
\frac {\d}{\d \bar\mu_{\bar j_1}^{\bar \b_1}}\ldots
\frac {\d}{\d \bar\mu_{\bar j_p}^{\bar \b_p}}\0\\
&\cdot& {\rm exp} \Big( - \chi^T {\cal K}^{-1} M - \frac 12
M^T {\cal V}{\cal K}^{-1}M\Big)
|\ES_\perp\rangle)\Big\vert_{\mu=\bar \mu=0}\label{expeq}
\ee
where
\be
{\cal K} = \I -\T\X,\quad\quad
{\cal V} = \left(\matrix{{\V}^{11}& {\V}^{12}\cr
{\V}^{21}& {\V}^{22}\cr}\right)\label{KV}
\ee
and
\be
M= \left(\matrix{\mu \cr \bar \mu}\right),\quad\quad
\chi^T = (a^\dagger \V^{12},a^\dagger \V^{21}), \quad\quad
\chi^T {\cal K}^{-1} M = a^\dagger C' (\rho_1 \mu + \rho_2 \bar \mu)
\label{Mchi}
\ee
The explicit computation, at first sight, looks daunting.
However, we may avail ourselves of the following identities
\be
&&\xi^T ({\cal V}{\cal K}^{-1})_{\a\a}\zeta =
\xi^T \tau C'({\cal V}{\cal K}^{-1})_{\a\a}\tau C'\zeta =
\xi^T C' \frac {\T} {\I - \T^2} \zeta =0\0\\
&& \xi^T \tau C'({\cal V}{\cal K}^{-1})_{\a\a}\zeta =
\xi^T ({\cal V}{\cal K}^{-1})_{\a\a}\tau C'\zeta =
\xi^T \tau \frac {\T} {\I - \T^2} \zeta=-\kappa\label{res1}
\ee
for $\a = 1,2$, and
\be
&&\xi^T ({\cal V}{\cal K}^{-1})_{12}\zeta =
\xi^T \tau C'({\cal V}{\cal K}^{-1})_{21}\tau C'\zeta =
-\xi^T C' \frac {\T} {\I - \T^2} \zeta =0\0\\
&&\xi^T ({\cal V}{\cal K}^{-1})_{21}\zeta =
\xi^T \tau C'({\cal V}{\cal K}^{-1})_{12}\tau C'\zeta =
\xi^T C' \frac 1 {\I - \T^2} \zeta =0\0\\
&&\xi^T ({\cal V}{\cal K}^{-1})_{12}\tau C'\zeta =
\xi^T \tau C'({\cal V}{\cal K}^{-1})_{21}\zeta =
\xi^T \tau \frac 1{\I - \T^2} \zeta =-1\0\\
&&\xi^T \tau C'({\cal V}{\cal K}^{-1})_{12}\zeta =
\xi^T ({\cal V}{\cal K}^{-1})_{21}\tau C'\zeta =
-\xi^T \tau \frac {\T} {\I - \T^2} \zeta =\kappa\label{res2}
\ee
Moreover
\be
&&(\chi^T {\cal K}^{-1})_1\xi =0,\quad\quad
(\chi^T {\cal K}^{-1})_1\tau C'\xi= a^\dagger \tau\xi\0\\
&&(\chi^T {\cal K}^{-1})_2\xi =a^\dagger C'\xi,\quad\quad
(\chi^T {\cal K}^{-1})_2\tau C'\xi= 0 \label{res3}
\ee
with analogous equations for $\zeta$.

In evaluating (\ref{res1}, \ref{res2}, \ref{res3}) we have used
the methods of ref.\cite{RSZ3} (see also \cite{BMS2}), together with
eqs.(\ref{xizeta}, \ref{cond}). These results are all we need to
explicitly compute (\ref{expeq}). In fact it is easy to verify that
the latter can be mapped to a rather simple combinatorial problem.
To show this we introduce generic variables
$x,y, \bar x, \bar y$, and make the following formal replacements:
\be
&&A \equiv \chi^T {\cal K}^{-1} M \longrightarrow x (a^\dagger \tau \xi)
+ \bar y (a^\dagger C' \zeta),\0\\
&& B \equiv M^T {\cal V}{\cal K}^{-1}M
\longrightarrow (-\kappa xy +\kappa x \bar y -\bar x y -\kappa \bar x \bar y)\label{repl1}
\ee
and
\be
(\tau C' \xi)^\a_l\frac {\d}{\d \mu_{l}^{\a}}= \d_x,\quad
\zeta^\b_j\frac {\d}{\d \mu_{j}^{\b}}= \d_y, \quad\quad
(\tau C' \xi)^{\bar \a}_{\bar l}
\frac {\d}{\d {\bar\mu}_{\bar l}^{\bar \a}}= \d_{\bar x},\quad\quad 
\zeta^{\bar \b}_{\bar j}\frac {\d}{\d {\bar \mu}_{\bar j}^{\bar\b}}= \d_{\bar y},\label{repl2}
\ee
Then (\ref{expeq}) is equivalent to
\be
\d_x^k \d_y^k \d_{\bar x}^p \d_{\bar y}^p \left. e^{-A-\frac 12 B}
\right\vert_{x=\bar x = y = \bar y =0}\label{repl3}
\ee
This in turn can be easily calculated and gives
\be
\sum_{m=0}^{[p,k]} \x^m \frac {k!p!}{m!} 
\sum_{l=m}^{[p,k]} (-1)^{l+m} \left(\matrix {k \cr l}\right)
\left(\matrix {p \cr l}\right)\left(\matrix {l\cr m}\right)
\kappa^{p+k-l-2m}
\label{repl4}
\ee
where $[n,m]$ stands for the minimum between $n$ and $m$. 
Now we insert this back into the original equation 
(\ref{expanstar}), we find  
\be
&&|\Lambda_n\rangle * |\Lambda_{n'}\rangle = 
\sum_{k=0}^n \sum_{p=0}^{n'} \sum_{m=0}^{[p,k]} 
\sum_{l=0}^{[p-m,k-m]} \frac{(-1)^{p+k+l}}{m!}\label{interm}\\
&& \cdot \kappa^{n+n'-l-2m}
\left(\matrix {n\cr k}\right) \left(\matrix {n' \cr p}\right)
\left(\matrix {k \cr m}\right)\left(\matrix {k-m \cr l}\right)
\left(\matrix {p \cr l+m}\right)\, \x^m |\ES_\perp\rangle\0
\ee

In order to evaluate these summations we split them as follows
\be
\sum_{k=0}^n \sum_{p=0}^{n'} \sum_{m=0}^{[p,k]}
\sum_{l=0}^{[p-m,k-m]} (\ldots) =
\sum_{k=0}^n\left(\sum_{p=k+1}^{n'}\sum_{m=0}^k\sum_{l=0}^{k-m} +
\sum_{p=0}^k \sum_{m=0}^p \sum_{l=0}^{p-m}\right)(\ldots)\label{a1}
\ee
Next we replace $l \to l+m$ and (\ref{a1}) becomes
\be
&&\sum_{k=0}^n\left(\sum_{m=0}^k\sum_{l=m}^k \sum_{p=k+1}^{n'}  +
\sum_{m=0}^k \sum_{p=m}^k \sum_{l=m}^{p}\right)(\ldots) = \0\\
&&=\sum_{k=0}^n\left(\sum_{m=0}^k\sum_{l=m}^k \sum_{p=k+1}^{n'}  +
\sum_{m=0}^k \sum_{l=m}^k \sum_{p=l}^{k}\right)(\ldots) =
\sum_{k=0}^n \sum_{m=0}^k \sum_{l=m}^k \sum_{p=l}^{n'}(\ldots)\label{a2}
\ee
Summarizing, we have now to calculate
\be
 |\Lambda_n\rangle * |\Lambda_{n'}\rangle &=&
\sum_{k=0}^n \sum_{m=0}^k \sum_{l=m}^k \sum_{p=l}^{n'}
\frac{(-1)^{p+k+l+m}}{m!}\0\\
&& \cdot \kappa^{n+n'-l-m}
\left(\matrix {n\cr k}\right) \left(\matrix {n' \cr p}\right)
\left(\matrix {k \cr m}\right)\left(\matrix {k-m \cr l-m}\right)
\left(\matrix {p \cr l}\right)\, \x^m |\ES_\perp\rangle\label{a3}
\ee
Now
\be
\sum_{p=l}^{n'} (-1)^{p+l} \left(\matrix {n' \cr p}\right)
\left(\matrix {p \cr l}\right) = \left(\matrix {n' \cr l}\right)
\sum_{p=0}^{n'-l} (-1)^{p} \left(\matrix {n'-l \cr p}\right) =
\left(\matrix {n' \cr l}\right) (1-1)^{n'-l}\label{a4}
\ee
This vanishes unless $l=n'$. In the case $n' >n$, $l<n'$.
Inserting this into (\ref{a3}), for $n'>n$ we get 0.

In the case $n=n'$, $l$ can take the value $n'$. This corresponds to
the case $k=p=l=n=n'$ in eq.(\ref{a3}). The result is easily derived
\be
|\Lambda_n\rangle * |\Lambda_{n}\rangle =\sum_{m=0}^n \frac{(-1)^{n+m}}
{m!}  \left(\matrix {n \cr m}\right)\kappa^{n-m}\x^m |\ES_\perp\rangle =
(-\kappa)^n L_n\left(\frac {\x}{\kappa} \right)|\ES_\perp\rangle = |\Lambda_n\rangle
\label{a5}
\ee
This proves eq.(\ref{nstarm}).

One could as well derive these results numerically. For instance, in order to
obtain (\ref{a5}) one could proceed, alternatively, as follows.
After setting $n=n'$ in (\ref{interm}), one realizes
that $|\Lambda_n\rangle * |\Lambda_{n}\rangle$ has the form
\be
|\Lambda_n\rangle * |\Lambda_{n}\rangle= \sum_{m=0}^n
F_m^{(n)}\left(\frac{\x}{\kappa}\right)^m|\ES_\perp\rangle\label{A7}
\ee
where
\be
F_m^{(n)} &=& 2\sum_{p=0}^{n-m}\sum_{k=0}^p \sum_{l=0}^k
\frac {(-1)^{p+k+l} \kappa^{2n-l-m}(n!)^2}{(m!)^2 (n-k-m)! (n-p-m)! l! (l+m)!
(k-l)!(p-l)!}\0\\
&& -\sum_{p=0}^{n-m}\sum_{l=0}^p
\frac {(-1)^{l} \kappa^{2n-l-m}(n!)^2}{[m!(n-p-m)!(p-l)!]^2 l!l+m)!}
\label{A8}
\ee
This corresponds to the desired result if
\be
F_m^{(n)}= \frac {(-1)^{n+m}}{m!} \kappa^n\left(\matrix {n \cr m}\right)
\label{A9}
\ee
Using {\it Mathematica} one can prove (numerically) that this is true
for any value of $n$ and $m$ a computer is able to
calculate in a reasonable time.

\subsection{Proof of eq.(1.3)}

The value of the SFT action for any solution $|\Lambda_n\rangle$  
is given by 
\be
{\cal S}(\Lambda_n) = {\EuScript K} 
\langle\Lambda_n|\Lambda_n\rangle\label{action2}
\ee
where ${\EuScript K}$ contains the ghost contribution. As shown 
in \cite{GRSZ1}, ${\EuScript K}$  is infinite unless it is suitably 
regularized. Nevertheless, as argued there, 
$|\Lambda_n\rangle$, together with the corresponding ghost solution,
can be taken as a representative of a corresponding class 
of smooth solutions. 

Our task now is to calculate $\langle\Lambda_n|\Lambda_n\rangle$.
However it may be important to consider states which are
linear combinations of $|\Lambda_n\rangle$. In order to evaluate their
action we have to be able to compute $\langle\Lambda_n|\Lambda_{n'}\rangle$.
Without loss of generality we can assume $n'>n$.
By defining $\tilde \x =(a^\dagger \tau C'\xi)\, (a^\dagger \zeta)$ we get
\be
\langle\Lambda_n|\Lambda_{n'}\rangle \!&=&\! (-\kappa)^{n+n'}\langle \tilde 0|
L_n(\tilde \x /\kappa)e^{-\frac 12 a \tilde \ES a} L_{n'} 
(\x / \kappa)
e^{\frac 12 a^\dagger \ES a^\dagger} |\tilde 0 \rangle\0\\
\!&=&\! L_n\left(\frac 1{\kappa}(\tau C' \xi)^\a_l \zeta^\b_j
\frac {\d}{\d \lambda_{l}^{\a}}\frac{\d}{\d \lambda_{j}^{\b}}\right)
L_{n'} \left(\frac 1{\kappa} (\tau \xi)^{ \a}_{l}(C'\zeta)^{\b}_{j}
\frac {\d}{\d {\mu}_{l}^{\a}} \,
\frac {\d}{\d {\mu}_{j}^{\b}}\right)\label{interm2} \0\\
\!&\cdot &\! \frac 1{\sqrt{\det (\I- \T^2)}}
\left. e^{\lambda C' \frac 1{ \I- \T^2}C'\mu
-\frac 12 \lambda C' \frac {\T}{\I- \T^2}\lambda
-\frac 12 \mu \frac {\T}{\I- \T^2}C'\mu}\right\vert_{\lambda=\mu=0}
\ee
For the derivation of this equation, see \cite{RSZ2,KP,RSZ3}.
Now, let us set
\be
A = \lambda C' \frac 1{\I- \T^2}C'\mu, \quad\quad
B= \lambda C' \frac {\T}{\I- \T^2}\lambda, \quad\quad
C = \mu \frac {\T}{\I- \T^2}C'\mu\0
\ee
and introduce the symbolic notation
\be
(\tau C' \xi)^\a_l\frac {\d}{\d \lambda_{l}^{\a}}= \d_x,\quad
\zeta^\b_j\frac {\d}{\d \lambda_{j}^{\b}}= \d_y, \quad\quad
(\tau \xi)^{\a}_{ l}
\frac {\d}{\d \mu_{l}^{\a}}= \d_{\bar x},\quad\quad
(C'\zeta)^{\b}_{ j}\frac {\d}{\d \mu_{j}^{\b}}=
\d_{\bar y},\label{repl2'}
\ee
Then, using (\ref{cond}) and (\ref{res1}, \ref{res2}), we find
\be
&&\d_x\d_{\bar x} A =0,\quad\quad \d_x\d_{\bar y} A =-1,
\quad\quad \d_y\d_{\bar x} A =-1,\quad\quad \d_y\d_{\bar y} A =0
\0\\
&&\d_x\d_{x} B =0,\quad\quad \d_x\d_{y} B =-2\kappa,\quad\quad
\d_y\d_{y} B =0,\label{repl3'}\\
&&\d_{\bar x}\d_{\bar x}C =0 ,\quad\quad \d_{\bar x}\d_{\bar y}C =-2\kappa,
\quad\quad \d_{\bar y}\d_{\bar y}C = 0\0
\ee
We can therefore make the replacement
\be
A-\frac 12 B - \frac 12 C \rightarrow \kappa
xy +\kappa \bar x\bar y - x \bar y - \bar x y\label{repl4'}
\ee
In (\ref{interm2}) we have to evaluate such terms as
\be
\d_x^k \d_y^k \d_{\bar x}^p\d_{\bar y}^p
(\kappa xy +\kappa \bar x\bar y - x \bar y - \bar x y)^{k+p}\0
\ee
for any two natural numbers $k$ and $p$. It is easy to obtain
\be
\frac 1{(p+k)!}\d_x^k \d_y^k \d_{\bar x}^p\d_{\bar y}^p
(\kappa xy +\kappa \bar x\bar y - x \bar y - \bar x y)^{k+p}=
\sum_{s=0}^{[p,k]}\left(\matrix{k \cr s}\right)
\left(\matrix{p \cr s}\right) \,k!\,p!\, \kappa^{p+k-2s}
\label{interm3}
\ee
Therefore we have
\be
&&\langle\Lambda_n|\Lambda_{n'}\rangle\0\\
\!&=&\! \sum_{k=0}^n \sum_{p=0}^{n'}
\frac {(-1)^{k+p}\kappa^{n+n'-p-k}}{k!p!} \left(\matrix{n \cr k}\right)
 \left(\matrix{n' \cr p}\right)
\d_x^k \d_y^k \d_{\bar x}^p\d_{\bar y}^p \left.
e^{A-\frac 12 B-\frac 12 C} \right\vert_{x=y=\bar x=\bar y=0}
\langle \ES_\perp|\ES_\perp\rangle \0\\
\!&=&\! \sum_{k=0}^n \sum_{p=0}^{n'}
\frac {(-1)^{k+p}}{k!p!} \left(\matrix{n \cr k}\right)
 \left(\matrix{n' \cr p}\right) \,\sum_{s=0}^{[p,k]}
\left(\matrix{k \cr s}\right)
\left(\matrix{p \cr s}\right) k!p! \kappa^{n+n'-2s}
\langle \ES_\perp|\ES_\perp\rangle \label{nn}
\ee
As in the previous subsection, we can rearrange the summations
as follows,
\be
&&\sum_{k=0}^n\sum_{p=0}^{n'} \sum_{p=0}^{[p,k]} (\ldots) =
 \sum_{k=0}^n \left( \sum_{p=0}^k \sum_{s=0}^p +
\sum_{p=k+1}^{n'} \sum_{s=0}^k\right) (\ldots)\label{b1}\\
&&=\sum_{k=0}^n \left( \sum_{s=0}^k \sum_{p=s}^k +
\sum_{s=0}^k\sum_{p=k+1}^{n'}\right)
(\dots)= \sum_{k=0}^n \sum_{s=0}^k \sum_{p=s}^{n'}(\ldots)\0
\ee
In conclusion we have to compute
\be
\langle\Lambda_n|\Lambda_{n'}\rangle =\sum_{k=0}^n \sum_{s=0}^k
\sum_{p=s}^{n'}\frac{(-1)^{p+k}\,n! n'!}
{(n-k)!(n'-p)! (k-s)!(p-s)! (s!)^2}\,\kappa^{n+n'-2s}
\langle \ES_\perp|\ES_\perp\rangle 
\label{b2}
\ee

Now,
\be
 \sum_{p=s}^{n'}(-1)^p \frac 1{(n'-p)!(p-s)!} =
 \sum_{p=0}^{n'-s}\frac {(-1)^{p+s}}{(n'-s)!}\left(\matrix{n'-s\cr p}\right)=
 \frac {(-1)^s}{(n'-s)!} (1-1)^{n'-s}\label{b3}
\ee
The right end side vanishes if $n' \neq s$, which is certainly true if $n'>n$. Therefore
in such a case, inserting (\ref{b3}) into (\ref{b2}) we get
$\langle\Lambda_n|\Lambda_{n'}\rangle =0$. When $s=n'$, eq.(\ref{b3}) is ambiguous. But this
corresponds to $p=k=s=n=n'$ in (\ref{b2}). The relevant contribution is elementary to compute, and one gets
\be
\langle\Lambda_n|\Lambda_n\rangle =
\langle\Lambda_0|\Lambda_0\rangle   \label{nn1}
\ee
This completes the proof of (\ref{nm}).

\section{The field theory limit and the GMS solitons}

In \cite{BMS2} we calculated the low energy limit of
$|\Lambda_0\rangle$. This is the Seiberg--Witten limit, 
\cite{SW}, defined by $\a' \rightarrow 0$ with $\alpha' B \gg g$, 
where $g$ is the closed 
string metric, in such a way that $G$, $\theta$ and $B$ are kept 
fixed. It was shown in \cite{BMS2} that
the three string vertex (\ref{VVmn}) becomes
\be
\V_{00}^{\a\b,rs} & \rightarrow &\, 
G^{\a\b}\delta^{rs}- \frac {4}{4a^2+3}
\left(G^{\a\b} \phi^{rs} -ia \epsilon^{\a\b}\chi^{rs}
\right)\\
\V_{0n}^{\a\b,rs} & \rightarrow &\, 0\\
\V_{mn}^{\a\b,rs} & \rightarrow &\, G^{\a\b}V_{mn}^{rs}\label{VVmnt}
\ee
so that the lump state factorizes into two factors, the first involves
only the zero modes, while the second contains only non--zero modes.
In particular we have
\be
&&\ES_{00}^{\a\b}=\frac{2|a|-1}{2|a|+1}\, G^{\a\b} \equiv s\,
G^{\a\b} \label{expB}
\ee
where $a$ has been defined above, (\ref{definitions}).

Finally, in this limit, the lump state $|\Lambda_0\rangle \equiv
|\ES\rangle$ $\to$ $|\hat \ES\rangle$, where
\be
|\hat \ES\rangle \!&=&\! \left\{{\rm det}(1-X)^{1/2}{\rm det} (1+T)^{1/2}\right\}^{26}
{\rm exp}\left(-\frac 12 \eta_{ \mu \nu}\sum_{m,n\geq 1}
a_m^{ \mu\dagger}S_{mn}a_n^{ \nu\dagger}\right)|0\rangle \otimes
\label{solB}\\
& & \frac{4a}{2a+1}\,\, \frac{b^2}{\sqrt{2\pi b^3}(\det G)^{1/4}}\,\,
{\rm exp}\left(-\frac 12
s
a_0^{\a\dagger}G_{\a\b}a_0^{\b\dagger}\right)| \Omega_{b, \theta}\rangle,\0
\ee
where  $\mu, \nu = 0, \dots 25$ and $\a, \b = 24, 25$
\footnote{The $a_n^{\a}, a_n^{\a\dagger}$ operators must be suitably rescaled
in order to absorb the metric factor in the exponent
$a_n^{\a\dagger}G_{\a\b}S_{nm} a_m^{\b\dagger}$ of the squeezed
state so that it takes the form appropriate for the sliver.}. 
The norm of the
lump is now regularized by the presence of $a$ which is proportional
to $B$, eq.(\ref{definitions}).
Using
\be
&& | x \rangle = \sqrt{\frac{2\sqrt{\det G}}{b\pi}}
\exp\left[-\frac{1}{b}x^{\a}G_{\a\b}x^{\b}
-\frac{2}{\sqrt{b}}i a_0^{\a\dagger} G_{\a\b}x^{\b}
+\frac{1}{2}a_0^{\a\dagger}G_{\a\b}a_0^{\b\dagger}
\right]
|\Omega_{b, \theta} \rangle
\ee
we can calculate the projection onto the basis of position
eigenstates of the transverse part of the lump state
\be
\langle x | e^{-\frac{s}{2}a_0^{\a \dagger}G_{\a\b}a_0^{\b \dagger}}|\Omega_{b, \theta} \rangle & = &
\sqrt{\frac{2\sqrt{\det G}}{b\pi}} \frac{1}{1+s}\,
 e^{-\frac{1-s}{1+s}\frac{1}{b}x^{\a}x^{\b}G_{\a\b}} \label{xproj}
\ee
Finally, the lump state projected into the $x$ representation is
\be
\langle x |\hat\ES\rangle
= \frac{1}{\pi}\, \exp \Big[-\frac{1}{2|a|b}x^{\a}x^{\b}G_{\a\b}\Big]|\Xi\rangle
=  \frac{1}{\pi}\, \exp \Big[-\frac{x^{\a}x^{\b}\delta_{\a\b}}{\theta}
\Big]
|\Xi\rangle \label{regul}
\ee
$|\Xi\rangle$ is the sliver state (RHS of first line in eq.(\ref{solB}))
and $\theta = \frac 1B$. We recall that $B$ has been chosen
nonnegative.

In order to analyze the same limit for any $|\Lambda_n\rangle$,
first of all we have to find the low energy limit of the
projectors $\rho_1, \rho_2$. Also these two projectors factorize
into the zero mode and non--zero mode part. The former is given by
\be
(\rho_1)_{00}^{\a\b}\rightarrow \frac 12 \Big[G^{\a\b}+ i \epsilon^{\a\b}\Big],
\quad\quad
(\rho_2)_{00}^{\a\b}\rightarrow \frac 12 \Big[G^{\a\b}- 
i \epsilon^{\a\b}\Big],\label{rholimit}
\ee

Now, in order to single out the appropriate limit of
$|\Lambda_n\rangle$, we take, in the definition (\ref{x}),
$\xi= \hat \xi+ \bar\xi$ and 
$\zeta= \hat \zeta + \bar\zeta$, where 
$\bar\xi, \bar\zeta$ vanish in the
limit $\a' \to 0$. Then we make the choice  
$\hat\xi_n = \hat \zeta_n = 0, \,\,\, \forall n>0$ and determine
$\hat \xi$ and $\hat \zeta$ in such a way that  
eqs.(\ref{xizeta}, \ref{hermit}) and (\ref{cond}) are satisfied in the 
limit $\a'\to 0$ (a detailed discussion of this limit
is contained in Appendix B). In the field theory limit the conditions 
(\ref{xizeta}) become
\be
\hat\xi_{0,24} + i \hat\xi_{0,25}=0, \quad\quad \hat\zeta_{0,24} 
+ i \hat\zeta_{0,25}=0,
\label{0cond}
\ee
From now on we set $\hat\xi_0 = \hat\xi_{0,25}= - i \hat\xi_{0,24}$ 
and, similarly,
$\hat\zeta_0 = \hat\zeta_{0,25}= - i\hat \zeta_{0,24}$.
The conditions (\ref{cond}) become
\be
&&\xi^T\tau \frac 1{\I-\T^2}\zeta \rightarrow - \frac{1}{1-s^2}
\frac 2{\sqrt{\det G}}{\hat\xi_0\hat\zeta_0} =-1\label{cond1}\\
&&\xi^T\tau \frac {\T}{\I-\T^2}\zeta \rightarrow -\frac{s}{1-s^2}
\frac 2{\sqrt{\det G}}{\hat\xi_0\hat\zeta_0}  =-\kappa\label{cond2}
\ee
Compatibility requires
\be
\frac{2\hat\xi_0\hat\zeta_0}{\sqrt{\det G}}= 1-s^2,\quad\quad \kappa= s
\label{comp}
\ee
At the same time
\be
(\xi \tau a^\dagger)(\zeta C' a^\dagger) \rightarrow -  
{\hat\xi_0\hat\zeta_0}
((a_0^{24\dagger})^2+(a_0^{25\dagger})^2)= -\frac {\hat\xi_0\hat\zeta_0}
{\sqrt{\det G}} a_0^{\a\dagger} G_{\a\b}a_0^{\b\dagger}\label{xitaua}
\ee
Hermiticity requires that the product $\hat\xi_0
\hat\zeta_0$ be real, in accordance with (\ref{cond1},\ref{cond2}).
Henceforth we will refer to the solutions found in this way
as the {\it factorized solutions}, since, as will become clear
in a moment, they realize the factorization of the star product
into the Moyal product and Witten's * product.
In order to be able to compute $\langle x| \Lambda_n\rangle$ in the
field theory limit, we have to evaluate first
\be
\langle x | \left(a_0^{\a\dagger} G_{\a\b}a_0^{\b\dagger}\right)^k\,
e^{-\frac{s}{2} a_0^{\a\dagger} G_{\a\b}a_0^{\b\dagger} }
|\Omega_{b, \theta} \rangle &=&
(-2)^k \frac {d^k}{d s^k} \left( \langle x |
e^{-\frac{s}{2} a_0^{\a\dagger}
G_{\a\b}a_0^{\b\dagger} }|\Omega_{b, \theta} \rangle\right)
\label{xkproj}\\
&=& (-2)^k \frac {d ^k}{d s^k} \left( \sqrt{\frac{2\sqrt{\det G}}{b\pi}}
\frac{1}{1+s}\,
 e^{-\frac{1-s}{1+s}\frac{1}{b}x^{\a}G_{\a\b}x^{\b}}\right) \0
\ee
An explicit calculation gives
\be
&& \frac {d^k}{d s^k} \left(\frac{1}{1+s}\,
 e^{-\frac{1-s}{1+s}\frac{1}{b}x^{\a}x^{\b}G_{\a\b}}\right) =\label{dsk}\\
&&\quad\quad= \sum_{l=0}^k \sum_{j=0}^{k-l}
\frac {(-1)^{k+j}}{(1-s)^j(1+s)^{k+1}}
\frac{k!}{j!} \left(\matrix{k-l-1\cr j-1}\right)\, \langle x,x \rangle^j
e^{-\frac{1}{2}\langle x,x\rangle}\0
\ee
where we have set
\be
\langle x,x \rangle = \frac{1}{ab}x^{\a} G_{\a\b}x^{\b} =
\frac{ 2 r^2 }{\theta}
\ee
with $r^2 = x^\a x^\b \delta_{\a\b}$.
In this equation
it must be understood that, by definition, the binomial coefficient
\mbox{ \small $\left(\matrix{-1\cr -1}\right)$} equals 1.

Now, inserting (\ref{dsk}) in the definition of $|\Lambda_n\rangle$,
we obtain after suitably reshuffling the indices:
\be
 \langle x| (-\kappa)^n L_n\Big(\frac{\x}{\kappa}\Big)
e^{-\frac 12 s a^{\a \dagger}_0 G_{\a\b} a^{\b \dagger}_0}
|\Omega_{b, \theta}\rangle
 & \to & \langle x | (-s)^n L_n \Big(-\frac{1-s^2}{2s}\,\,
a_{0}^{\a\dagger}G_{\a\b}a_{0}^{\b\dagger}\Big)
e^{-\frac 12 s a_0^{\a \dagger} G_{\a\b} a_0^{\b \dagger}}|\Omega_{b, \theta}\rangle \0\\
&=&
\frac{(-s)^n}{(1+s)} \sum_{j=0}^n\sum_{k=j}^n \sum_{l=j}^k
\left(\matrix{n\cr k}\right)\left(\matrix{l-1\cr j-1}\right)\frac 1{j!}
\frac {(1-s)^k}{(1+s)^{j}s^k}\0\\
&\cdot& (-1)^j \langle x,x \rangle^j\,
e^{-\frac{1}{2}\langle x,x \rangle}
\sqrt{\frac{2\sqrt{\det G}}{b\pi}} \label{xlambda1}
\ee
The expression can be evaluated as follows. First one uses the result
\be
\sum_{l=j}^k \left(\matrix{l-1\cr j-1}\right) = \left(\matrix{k\cr j}\right) \label{c1}
\ee
Inserting this into (\ref{xlambda1}) one is left with the following summation, which contains an
evident binomial expansion,
\be
\sum_{k=j}^n   \left(\matrix{n\cr k}\right)\left(\matrix{k\cr j}\right) \left(\frac{1-s}{s}\right)^k=
\left(\matrix{n\cr j }\right) \frac{(1-s)^j}{s^n}\label{c3}
\ee
Replacing this result into (\ref{xlambda1}) we obtain
\be
\langle x| (-\kappa)^n L_n\left(\frac{\x}{\kappa}\right)
e^{-\frac 12 s a^{\a \dagger}_0 G_{\a\b} a^{\b \dagger}_0}|\Omega_{b, \theta}\rangle &\rightarrow&
\frac{2|a|+1}{4|a|}\sqrt{\frac {2\sqrt{\det G}}{b \pi }}(-1)^n \sum_{j=0}^n
\left(\matrix{n\cr j}\right)\frac 1{j!} \,
\left(- \frac{2r^2}{\theta}\right)^j
e^{-\frac {r^2}{\theta}}\0
\ee
Recalling now
that the definition of $|\hat \ES\rangle$ includes an additional
numerical factor (see eq.(\ref{solB})), we finally obtain
\be
\langle x| \Lambda_n\rangle \rightarrow \langle x| \hat\Lambda_n\rangle&=& \frac 1{\pi}(-1)^n \sum_{j=0}^n
\left(\matrix{n\cr j}\right)\frac 1{j!} \,
\left(- \frac{2r^2}{\theta}\right)^j
e^{-\frac {r^2}{\theta}} |\Xi\rangle \0\\
&=& \frac 1{\pi}(-1)^n\,L_n\left(\frac{2r^2}{\theta}\right) 
\,e^{-\frac {r^2}{\theta}}|\Xi\rangle
\ee
as announced in the introduction. The coefficient
in front of the sliver $|\Xi\rangle$ is the $n-th$ GMS
solution. Strictly speaking there is a discrepancy between
these coefficients and the corresponding GMS soliton, given
by the normalizations which differ by a factor of $2\pi$.
This can be traced back to the traditional normalizations 
used for the eigenstates $|x\rangle$ and $|p\rangle$ in the
SFT theory context and in the Moyal context, respectively.
This discrepancy can be easily dealt with with a simple redefinition.

Finally we remark that, for the solutions 
$\langle x| \hat\Lambda_n\rangle$, in view of the properties of the
GMS solitons, we have achieved factorization of the star product 
with $B$ field into the Moyal product and Witten's star product.
 
\section{Conclusion}

In \cite{GMS} it was shown that a generic noncommutative scalar
field theory with polynomial interaction allows for solitonic
solutions in any space dimension, see also \cite{Komaba} and references
therein. The solutions are very elegantly
constructed in terms of harmonic oscillators eigenstates
$|n\rangle$. In particular, solitonic solutions correspond
to projectors $P_n=|n\rangle\langle n|$. Via the Weyl transform these
projectors can be mapped to classical functions $\psi_n(x,y)$ of two variables
$x,y$, in such a way that the operator product in the Hilbert space
correspond to the Moyal product in $(x,y)$ space.

This construction is rather universal and does not depend in any
essential way on the form of the potential. Now, as we have noticed
in the introduction, the low energy effective tachyonic field
theory derived from SFT in the presence of a background $B$ field
is a noncommutative scalar field theory of the type described above.
Therefore it is endowed with the GMS noncommutative solitons.
It is reasonable to expect that these solitons may emerge from
soliton--type solutions of the SFT, which has the noncommutative
scalar tachyonic field theory as its low energy effective action.
Therefore the low energy GMS solitons we found in the
previous sections are no surprise. What is rewarding however
is the isomorphism we find between the lump solutions
$|\Lambda_n\rangle$ in VSFT and the corresponding GMS solitons.
Setting $r^2 = x^2+y^2$ and $\psi_n (x,y) =2(-1)^n\,
L_n(\frac{2r^2}{\theta}) \,e^{-\frac {r^2}{\theta}}$,
we have in fact the following correspondences
\be
\matrix{|\Lambda_n\rangle &\longleftrightarrow&
P_n
&\longleftrightarrow&\psi_n (x,y)  \cr
|\Lambda_n\rangle * |\Lambda_{n'}\rangle &\longleftrightarrow&
P_n P_{n'}&\longleftrightarrow &\psi_n \star \psi_{n'}}\label{c2}
\ee
where $\star$ denotes the Moyal product. Moreover
\be
\langle \Lambda_n|\Lambda_{n'}\rangle &\longleftrightarrow&
{\rm Tr} (P_nP_{n'}) \,\longleftrightarrow\,
\int dxdy \,\psi_n(x,y)\psi_{n'}(x,y)
\ee
up to normalization (see (\ref{nm})).
This correspondence seems to indicate that the Laguerre polynomials
hide a universal structure of these noncommutative algebras.

It is evident from the above that the GMS solitons are the low energy
remnants of corresponding D--branes in SFT. This explains many features
of the former: why, for instance, the energy of the soliton
given by $\sum_{k=0}^{n-1} |k\rangle \langle k|$ is  $n$ time the energy
of the soliton $ |0\rangle \langle 0|$; this is nothing but a low 
energy relic
of the same property for the tensions of the corresponding D--branes.

\section{Appendices}
\subsection{Appendix A}

In this appendix we discuss compatibility among 
eqs.(\ref{xizeta}, \ref{hermit}) and (\ref{cond}). In particular 
we would like to argue that not only they 
 admit solutions $\xi$ and $\zeta$ but that the latter are 
expected to have an infinite number of undetermined components. 
To start with, from \cite{RSZ3} we deduce 
that the $\rho$ projectors, in the absence of a $B$ field, halve
the number of $\xi$ and $\zeta$ components. This will remain
true in the presence of the deformation due to the 
$B$ field (at least for generic values of $B$, but there is no
evidence of `critical' values of $B$ (see \cite{BMS2} 
for exact calculations in this sense)). Second, the
equation (\ref{hermit}) relates $\xi^*,\zeta^*$ to $\xi,\zeta$.
Third, the eqs.(\ref{cond}) is a finite set of 
conditions. Therefore, in a generic situation, we expect that there
exist $\xi$ and $\zeta$ solutions to these equations with infinite
many indeterminate components. The only trouble could come from 
a mutual incompatibility of these equations. Let us discuss
this point more closely by first solving explicitly eq.(\ref{hermit}).
This equation has two simple solutions: (I)  $\zeta = \tau \xi^*$
and (II) $\xi = C'\xi^*,\, \zeta= C'\zeta^*$. Let us analyze their 
compatibility with the remaining equations. To do so we have to first
recall eqs.(\ref{X}, \ref{commute}, \ref{CV}) and (\ref{rho1}, \ref{rho2},
\ref{proj12}, \ref{rhorels}). From the properties quoted in section 3 
of \cite{BMS} one easily gets in addition 
\be
(\V^{rs})^* = \tilde\V^{rs}\label{V*V}
\ee
For clarity we recall that the $^*$ superscript denotes complex 
conjugation, $\,\tilde{}\,$ represents transposition with respect to the
$\a,\b$ indices, while $^T$ is the transposition with respect to
both $\a,\b$ and $N,M$ indices. We also introduce the operation 
$^\dagger \, =\, ^{*T}$. Using all this it is easy to prove that
\be
&&(\X^{rs})^* = \tilde \X^{rs}, \quad {\rm i.e.}\quad (\X^{rs})^\dagger=
\X^{rs}\0\\
&&[(\X^{rs})^*, (\X^{r's'})^*]=0,\quad\quad  
[\tilde \X^{rs}, \tilde \X^{r's'}]=0,\label{Xrhoprop}\\
&&\rho_i^\dagger = \rho_i, \quad {\rm i.e.}\quad \rho_i^*= \tilde\rho_i,
\quad i=1,2\0\\
&&\tau \rho_i = \tilde \rho_i \tau, \quad \quad i=1,2\0
\ee
Now we are ready to verify compatibility of (I) and (II) with 
eqs.(\ref{xizeta}, \ref{cond}). 
As for the former we have
\be
\rho_1\zeta = \rho_1 \tau \xi^*= \tau \tilde \rho_1 \xi^*= \tau
\rho_1^*\xi^* = \tau (\rho_1\xi)^*=0\label{comp1}
\ee
Therefore (I) is consistent with (\ref{xizeta}). In case (II)
we have instead
\be
\rho_1 \xi = \rho_1 C'\xi^* = C'\tilde \rho_2\xi^*= C'\rho_2^*\xi^*
=C'\xi^*\label{comp2}
\ee
so that requiring $\rho_1\xi=0$ implies $\xi=0$ identically.
Therefore we must discard solution (II).

Next we have to prove compatibility of (I) with (\ref{cond}), in other
words we must show that when replacing (I) into the RHS of the
two eqs.(\ref{cond}) we get real numbers. Let us show this for the first
equation, because for the second no significant modification is needed
\be
\left(\xi^T \tau\frac 1{\I-\T^2}\zeta\right)^*= 
\xi^{T*} \tau\frac 1{\I-(\T^*)^2}\zeta^* =
\zeta^T \frac 1{\I-(\T^*)^2}\tau\xi = 
\xi^T \tau\frac 1{\I-(\T^\dagger)^2}\zeta=
\xi^T \tau\frac 1{\I-\T^2}\zeta \0
\ee
where the second equality is obtained by replacement of (I), and the 
third by transposition.    

To complete our argument we have to prove compatibility of 
eq.(\ref{xizeta}) with (\ref{cond}). Let us impose that $\xi$ and $\zeta$ 
are solutions to (\ref{xizeta}),
\be
\xi^T \tau\frac 1{\I-\T^2}\zeta = 
\xi^T \tau\frac 1{\I-\T^2}\rho_2\zeta = 
\xi^T \tau\rho_2\frac 1{\I-\T^2}\zeta=
\xi^T\tilde \rho_2 \tau\frac 1{\I-\T^2}\zeta =
\xi^T \tau\frac 1{\I-\T^2}\zeta \label{verif2}
\ee
where the second equality is a consequence of $[\rho_i, \T]=0$,
while the last follows from
\be
\xi^T= (\rho_2\xi)^T = \xi^T \rho_2^T =\xi^T\tilde\rho_2\0
\ee
We see that no additional constraints are obtained.

It remains for us to examine the question of whether the solution 
(\ref{unicum})
to (\ref{hermit}) is the only possible one. We do not have a formal proof of 
this fact, but we can produce the following argument. On the LHS of eq.(\ref{hermit})
the vectors $\tau\xi^*$ and $C'\zeta^*$ lie in the subspaces annihilated by
$\rho_1$ and $\rho_2$, respectively. Similarly on the RHS of the same equation
$\tau C'\xi$ and $\zeta$ lie in subspaces annihilated by 
$\rho_2$ and $\rho_1$,
respectively. Since the subspaces annihilated by $\rho_1$ and $\rho_2$
are linearly independent, the equality of the two sides of eq.(\ref{hermit})
should automatically imply eq.(\ref{unicum}) (up to an overall constant factor
which can be absorbed by rescaling $\zeta$ and $\xi$ in opposite directions)
\footnote{We thank one of the referees of this paper for suggesting this
argument.}.

\subsection{Appendix B}  

This appendix is devoted to a discussion of the factorized 
$\a'\to 0$ limit found after eq.(\ref{rholimit}).  There we wrote 
the vectors $\xi$ and $\zeta$ in the form $\xi= \hat \xi+ \bar\xi$ and 
$\zeta= \hat \zeta + \bar\zeta$, where 
$\bar\xi, \bar\zeta$ are supposed to vanish in the limit $\a' \to 0$. 
Then we made the choice $\hat\xi_n = \hat \zeta_n = 0, \,\,\, 
\forall n>0$ and determined
$\hat \xi$ and $\hat \zeta$ in such a way that  
eqs.(\ref{xizeta}, \ref{hermit}, \ref{cond}) be satisfied in the limit
$\a'\to 0$. We called such kind of solutions the factorized ones.
In section 4 it was tacitly assumed that such solutions are $\a'\to 0$ limits 
of $\a'\neq 0$ solutions. This is a reasonable assumption, based on
the fact that the $\a'\to 0$ limit is smooth in all the relevant 
equations (we mean specifically eqs.(\ref{xizeta}, \ref{cond})).
The latter are a system of algebraic (linear and quadratic) equations,
and we do not expect any kind of singularity in the $\a'\to 0$ limit, 
and, therefore, the factorized solutions should indeed be the limits 
of solutions to eqs.(\ref{xizeta}, \ref{hermit}, \ref{cond}) for 
generic $\a'$.

Here we would like to examine this question more closely. To confirm
our conjecture based on smoothness we will explicitly construct 
$\bar\xi$ and
$\bar\zeta$ as power series which satisfy 
eqs.(\ref{xizeta}, \ref{hermit}, \ref{cond}) and vanish for $\a'\to 0$.
We recall from \cite{MT} that the most appropriate way to define the
$\a'\to 0$ limit is to introduce a small dimensionless
parameter $\epsilon$ together with the rescalings
\be
V_{00}&\rightarrow& \e^2 V_{00}\0\\
V_{0n}&\rightarrow& \e V_{0n}\label{elimit}\\
V_{mn}&\rightarrow& V_{mn}\0
\ee
From now on we set $\a'$ to a constant value, say 1, and consider 
instead small $\e$ expansions. Let $f$ be a generic function of $\a'$, like
$\rho_i, \xi, \X,...$, and write it as
\be
f = \hat f+ \bar f = \hat f + \e f'+ \e^2 f'' + ...+ \e^n f^{(n)}+ ...
\label{expan}
\ee 
When convenient we will use the notation $f^{(0)}$ for $\hat f$.
We wish to satisfy eqs.(\ref{xizeta}, \ref{hermit}, \ref{cond}) 
in this $\e$--expanded form. As for (\ref{hermit}), we understand
that it is satisfied via (\ref{unicum}).
Next we deal with (\ref{xizeta}). Writing $\rho_1^2=\rho_1$ in
the expanded form, we find
\be
&&\hat \rho_1^2=\hat\rho_1 \0\\
&&\rho_1' =\hat\rho_1 \rho_1' + \rho_1'\hat\rho_1\0\\
&&\rho_1'' = {\rho_1'}^2 + \hat\rho_1\rho_1''+\rho_1''\hat\rho_1,\label{erho}\\
&&\ldots\0\\
&&\rho_1^{(n)} = \sum_{k=0}^n \, \rho_1^{(k)} \,\rho_1^{(n-k)}
\ee
Similarly, from $\rho_1\xi =0$, we get
\be
&&\hat\rho_1 \hat \xi = 0\label{e1}\\
&&\hat\rho_1 \xi' + \rho_1' \hat \xi =0\label{e2}\\
&&\rho_1'' \hat\xi + \rho_1'\xi' + \hat\rho_1 \xi''=0,\label{e3}\\
&&\ldots\0\\
&&\sum_{k=0}^n\, \rho_1^{(k)} \xi^{(n-k)}=0\label{2n}
\ee
A solution to (\ref{e1}) was found in section 4 (the factorized 
solution). If we replace the second of eqs.(\ref{erho}) into (\ref{e2})
we get $\hat\rho_1(\xi'+\rho_1'\hat\xi)=0$ which implies that
\be
\xi' = -\rho_1' \hat \xi +\eta'\label{xi'}
\ee
where $\eta'$ is any solution to the equation $\hat\rho_1\eta'=0$;
recalling Appendix A, $\eta'$ will have an infinite number of 
undetermined components. Analogously, for $\xi''$ we find
\be
\xi'' = -\rho_1'' \hat\xi - \rho_1'\xi' + \eta''\label{xi''}
\ee
where $\eta''$ is a new arbitrary solution to $\hat\rho_1\eta''=0$.
We remark that a term proportional to $\rho_1'{}^2 \hat \xi$
could be added to the RHS of eq.(\ref{xi''}). This follows from the fact 
that such a term is in the kernel of $\hat \rho_1$, for
\be
\hat \rho_1 \rho_1'\rho_1'\hat\xi = (\rho_1'\rho_1'- \rho_1'\hat\rho_1\rho_1')
\hat \xi= (\rho_1'(\rho_1'\hat\rho_1 +\hat\rho_1 \rho_1')- \rho_1'
\hat\rho_1\rho_1')\hat\xi=0\0
\ee
where we have used the relation $\rho_1' = \rho_1'\hat\rho_1+
\hat\rho_1 \rho_1'$ twice; the first time applied to the
product $\hat \rho_1 \rho_1'$ in the LHS, the second time in an 
obvious way. Therefore the addition of such a term simply amounts to a 
redefinition of $\eta''$.

Proceeding further as above, it is evident that at every new order of approximation for $\xi$
we get in addition a new arbitrary eigenvector of
$\hat\rho_1$ with eigenvalue 0. In general
\be
\xi^{(n)} = - \sum_{k=0}^{n-1} \rho_1^{(n-k)}\xi^{(k)} +\eta^{(n)}
\label{xin}
\ee
where $\eta^{(n)}$ is the eigenvector in question. 
It is possible to construct explicit examples of $\xi$ at the lowest orders in 
$\e$: for instance $\xi=\rho_2\hat\xi$ satisfies the above requirements to order zero and 1 in $\e$.

Now let us deal with (\ref{cond}). We first call
\be
F = \frac 1{\I-\T^2},\quad\quad G= \frac \T{\I-\T^2} \0
\ee
then, as above, expand the LHS of (\ref{cond}) in powers of $\e$.
We recall from section 4 that by construction
\be
\hat \xi \tau \hat F \hat\zeta = -1 ,\quad\quad 
\hat \xi \tau \hat G \hat\zeta = -\kappa\0
\ee 
Therefore the higher orders in the LHS of (\ref{cond}) must be equated
to 0. To first order in $\e$, after explicitly inserting (\ref{unicum}),
we get 
\be
\hat \xi \tau F' \tau \hat\xi^* -
\hat \xi \tau\hat F \tau{\rho_1'}^* \hat\xi^*-
\hat \xi{\rho_1'}^T \tau\hat F \tau \hat\xi^*+
\hat \xi \tau \hat F \tau{\eta'}^* +
\eta' \tau\hat F \tau \hat\xi^*=0\label{1o}
\ee
and an analogous equation with $F$ everywhere replaced by $G$.
In order to understand how these two equations can be satisfied we
have to go back to (\ref{elimit}). Inserting the latter in 
(\ref{VVmn}), one can see that the first order correction
to $\V_{00}^{rs}$ vanishes. The same holds for $\X_{00}^{rs}$, and,
due to the block diagonal form of the hatted objects 
$\hat \X^{rs}, \hat\rho_i,\hat F,...$, we come to the conclusion
that $(\rho_1')_{00}= (F')_{00}= (G')_{00}=0$ (in fact the first
nonvanishing corrections to the $M=0,N=0$ components come from
the second order in $\e$). Now recalling that the only nonvanishing
component of $\hat \xi$ is the zeroth one, we see that in (\ref{1o})
the first three terms vanish. Therefore (\ref{1o})
is satisfied provided we choose $\eta'_0 =0$. The same is true 
for the analogous equation with $F$ replaced by $G$.  

This proves that, to first order in $\e$, 
eqs.(\ref{xizeta}, \ref{hermit}, \ref{cond}) can be satisfied 
with $\xi' = \eta'$, where $\eta'$ has vanishing zeroth components,
while $\eta'_n$ are subject to the condition of defining an
eigenvectors of $\hat\rho_1$ with 0 eigenvalue, but are otherwise 
arbitrary. We remark that the precise form (\ref{elimit}) of the
rescaling plays a crucial role in satisfying (\ref{1o}) and 
its companion equation: were the first three terms of (\ref{1o})
nonvanishing, it would be impossible to satisfy both equations
for any value of $s$ because $\hat G_{00} = s \hat F_{00}$.

Let us analyze now the second order approximation in $\e$.
The equation analogous to (\ref{1o}) is 
\be
\xi'' \tau \hat F \hat\zeta+ \hat\xi \tau F'' \hat\zeta
+ \hat\xi \tau \hat F \zeta'' + \xi' \tau F' \hat \zeta
+ \xi' \tau \hat F \zeta' + \hat \xi \tau F' \zeta'=0\label{2o}
\ee
From the above discussion we know that in the first three terms
only the zeroth components of $\xi''$ and $\zeta''$ and
the $_{00}$ component of $F''$ do contribute. However here no
such simplification occurs as in (\ref{1o}), because all the 
components of $\xi',\zeta'=\tau{\xi'}^*$ and $F'$ will 
contribute (and the equality $\hat G_{00} = s \hat F_{00}$ is not 
dangerous in this
new context). But now we have at our disposal all the infinite
many arbitrary components of $\eta'$ to satisfy (\ref{2o}) and the 
companion equation with $F$ replaced by $G$. On the basis of the 
argument introduced in Appendix A, this is possible in an infinite 
number of ways.

Now the pattern is clear. At the third order in $\e$ we can count
on the infinite many arbitrary components of $\eta''$ to satisfy
the third order approximation of (\ref{cond}), and so on.
This completes our argument to show that the factorized 
solutions can be obtained as the $\a'\to 0$ limit of
$\a'\neq 0$ solutions of eqs.(\ref{xizeta}, \ref{hermit}, \ref{cond}).

\acknowledgments

L.B. would like to thank Branko Dragovich for pointing out to him 
ref.\cite{BD}, where GMS solitons were obtained in the framework of 
$p$--adic string theory. This research was supported by the Italian MIUR 
under the program ``Teoria dei Campi, Superstringhe e Gravit\`a''.

\end{document}